\documentclass[%
 aip,
cp,  
 amsmath,amssymb,
 reprint,%
]{revtex4-2}

\usepackage{graphicx}
\usepackage{dcolumn}
\usepackage{bm}

\usepackage[utf8]{inputenc}
\usepackage[T1]{fontenc}
\usepackage{mathptmx} 

\begin{document}

%
\title{The Shock Physics of Giant Impacts:\\ Key Requirements for the Equations of State}

\author{Sarah Stewart} 
 \email[Corresponding author: ]{sts@ucdavis.edu}
\author{Erik Davies}%
\author{Megan Duncan}%
\affiliation{U. California, Davis, CA 95616, USA}

\author{Simon Lock}
\affiliation{California Institute of Technology, Pasadena, CA 91125, USA}%

\author{Seth Root}
\author{Joshua Townsend}
\affiliation{Sandia National Laboratories, Albuquerque, NM 87185, USA}

\author{Richard Kraus}
\affiliation{Lawrence Livermore National Laboratory, Livermore, CA 94550, USA}

\author{Razvan Caracas}
\affiliation{National Center for Scientific Research, 69007 Lyon, France}

\author{Stein Jacobsen}
\affiliation{Harvard University, Cambridge, MA 02138, USA}

\date{\today} 

\begin{abstract}
We discuss major challenges in modeling giant impacts between planetary bodies, focusing on the equations of state (EOS). During the giant impact stage of planet formation, rocky planets are melted and partially vaporized. However, most EOS models fail to reproduce experimental constraints on the thermodynamic properties of the major minerals over the required phase space. Here, we present an updated version of the widely-used ANEOS model that includes a user-defined heat capacity limit in the thermal free energy term. Our revised model for forsterite (Mg$_2$SiO$_4$), a common proxy for the mantles of rocky planets, provides a better fit to material data over most of the phase space of giant impacts. We discuss the limitations of this model and the Tillotson equation of state, a commonly used alternative model.
\end{abstract}

\maketitle

\section{\label{sec:intro}MODELING GIANT IMPACTS BETWEEN PLANETS}

Rocky planets form by a series of giant impacts with sufficient energy to vaporize the outer layers of the bodies \cite[e.g.,][]{Canup2008,Nakajima2014,Lock2017}. In many giant impacts, the colliding bodies are transformed into a new type of astronomical object called a synestia \cite{Lock2017}, which is a body that exceeds the limit of a spheroidal shape. Most impact-generated synestias with an Earth-like composition have internal temperatures and pressures that reach supercritical conditions \cite{stewart2018raining}. Circumplanetary disks formed by giant impacts may be primarily liquid or primarily vapor, depending on the exact impact parameters. The escaping material is typically strongly shocked with a large vapor mass fraction. 

Accurately modeling a giant impact requires an equation of state (EOS) that can capture the wide range of shock-loading and release paths that occur during a single event. The importance of the EOS in hydrocode modeling is discussed in detail in \cite{melosh2007hydrocode}. Because much of the material will pass through or end up on a phase boundary, inclusion of realistic phase boundaries is essential. However, accurate wide-ranging EOS models are difficult to develop, and in simulations to date, the EOS models have substantial omissions or discrepancies with available experimental constraints on the material properties of major planetary constituents. Furthermore, giant impact simulations have typically simplified the materials in planets into  single-component layers. For rocky planets, most simulations have used a single-component silicate mantle and pure iron metal core. Future work will need to incorporate multi-component melting and vaporization into EOS models. 

Here, we present an updated version of the ANEOS model \cite{thompson1970improvements}, which is widely used by the planetary impact community. ANEOS is a set of equation of state routines developed for run-time calculations in hydrocodes. The model is formulated in terms of the Helmholtz free energies for solid, liquid, vapor, plasma and mixed phases. As a result, the model is capable of spanning nearly all pressures, densities, and temperatures encountered during natural phenomenon such as giant impacts. The model has been developed to increase flexibility in fitting the EOS \cite{thompson1974improvements,thompson1990aneos}, to include the presence of molecules in the gas \cite{melosh2007hydrocode}, and to enable the simultaneous inclusion of a melt curve and a high-pressure phase transition \cite{collins2014improvements}. In the ANEOS model, the thermal free energy for the solid is formulated using a Debye model with a classical Dulong-Petit limit. In the liquid region, the total Helmholtz free energy is the sum of the total solid Helmholtz free energy plus a contribution from melting (plus an electronic term at high temperatures). Thus, the same Debye thermal model is applied to both the solid and liquid phases. 

The specific heat capacities of silicate liquids are known to vary with composition and pressure. Based on calorimetry of silicate glasses and liquids at 1~bar \cite{stebbins1984heat,lesher2015thermodynamic}, the specific heat capacities of felsic liquids (rich in silica) are closer to the Dulong-Petit limit of $3nR$, where $n$ is the number of atoms per formula and $R$ is the gas constant. But the heat capacities of mafic compositions (rich in magnesium) are much larger. These differences at low pressures are primarily attributed to the higher degree of polymerization in felsic liquids than in mafic liquids. The specific heat capacity at constant volume of liquid forsterite is much larger than the Dulong-Petit limit at 1~bar ($c_v\sim4.2nR$ \cite{thomas2013direct}) and under shock compression ($4.2nR$ to $4.9nR$; values extracted from the density functional theory-based quantum molecular dynamics (QMD) calculations presented in \cite{Root2018}). At high pressures and temperatures, dissociation, electronic excitations, anharmonicity, and depolymerization are important processes \cite{Hicks2006}. Because ANEOS does not incorporate all of these processes, the temperatures and entropies on model Hugoniots deviate widely from experimental constraints \cite[e.g.,][]{Kraus2012,Davies2019foentropy}. In this work, we extend the ANEOS model to provide users with the ability to adjust the specific heat capacity limit in the thermal model to be able to fit available data, and we present a revised model for forsterite.

Finally, we discuss the limitations of our revised ANEOS model and the Tillotson equation of state, which is a commonly used alternative EOS for planetary collisions. The Tillotson model \cite{tillotson1962metallic} was developed to provide a pragmatic wide-ranging pressure-volume-energy formulation for metals. This simple analytic model interpolates between a condensed region and ideal monatomic gas region. The condensed region is based on the known shock Hugoniot using a linear particle velocity-shock velocity relation with a Mie-Gr\"uneisen thermal model that extrapolates to the Thomas-Fermi limit at high pressures. The model does not include any phase boundaries, and the interpolated region is not physically motivated. In some implementations, temperature is estimated assuming a constant specific heat capacity \cite[e.g.,][]{Brundage2013}. This analytic model is often used for computational expediency or when a more advanced material model is unavailable. Here, we illustrate why the Tillotson EOS is unable to model giant impacts where vaporization is an important physical process. 

\section{NEW USER-DEFINED SPECIFIC HEAT CAPACITY LIMIT IN THE ANEOS MODEL}

ANEOS produces a wide-ranging thermodynamically self-consistent EOS surface. However, the accuracy of the model can be very poor in certain regions of the phase diagram. In general, ANEOS uses relatively simple formulations for each region of phase space. More accurate EOS calculations are possible for individual phases but are rarely assembled into wide-ranging EOS that include the vapor and sublimation curves. One of the major issues in developing ANEOS models for geologic materials is the hard-coded Dulong-Petit limit. 

Giant impact applications typically involve warm planets, already near the solidus or partially molten. The peak shock pressures during terrestrial planet formation are typically 100’s of GPa, which deposits enough irreversible work to lead to widespread melting and partial vaporization. An ideal improvement to ANEOS would be the incorporation of a new liquid thermal model; however the Debye phonon model cannot be generalized easily for molecular liquids (e.g., \cite{bolmatov2012phonon}). Here, we implement an interim, practical modification to improve the ability to model giant impacts. Since the giant impact must capture the thermodynamics of the liquid region to be able to model the high-pressure regions of the shock Hugoniot and shock-produced melt, we multiply the thermal free energy term for the condensed phases by a user-defined value that is essentially a fitting parameter, $f_{cv}$. As a result, the specific heat capacity in the liquid region can be adjusted by the user. Note that deviating from the Dulong-Petit limit requires adjusting the reference Debye temperature and leads to limitations and errors in the solid region, which are discussed below.

An EOS model must span the regions of phase space between condensed phases and an ideal gas to be able to solve for a liquid-vapor curve via the Maxwell construction. The modified ANEOS uses the following equation for the thermal term in the Helmholtz free energy, $F_{th}$, to span this large region of phase space:
\begin{eqnarray}
    F_{th}(\rho,T) & = & N_0 k T \left \{ f_{cv} \left[ 3 \ln  \left(1 - e^{-\theta/T} \right)- \mathcal{D}(\theta / T) \right] + \frac{3}{2} \frac{1}{b} \ln \left( 1 + \psi^{b} \right) \right \}, {\rm where} \label{eq:interp} \\
    \psi(\rho,T) & = & Z(\rho,T) \frac{C_{13} \rho^{2/3} T}{\theta^2}, {\rm and}\\
    C_{13} & = & \frac{N_0^{5/3} h^2}{2 \pi k} \exp \left \{ \frac{2}{3} \sum_l \frac{N_l}{N_0} \ln \left( \frac{N_l}{N_0^{5/2} M_l^{3/2}} \right) \right \}.
\end{eqnarray}
$N_0$ is the average number of atoms per unit mass; $k$ is the Boltzmann constant; $T$ is temperature; $\theta$ is the density-dependent Debye temperature; $\mathcal{D}$ is the third order Debye function; $\rho$ is density; $Z(\rho,T)$ is the function that incorporates the free energy adjustment for molecules from \cite{melosh2007hydrocode}; $N_l$ is the number of atoms with atomic mass $M_l$; and $h$ is Planck's constant. $f_{cv}$ is the new term introduced here to modify the heat capacity limit to $3f_{cv}N_0kT$. The adjustment parameter $b$ is used to modify the critical point \cite{thompson1990aneos}. $\psi$ is a term that represents how closely the material corresponds to an ideal gas. When the temperatures are large and/or the densities are low, $\psi$ is large and the right hand term in Eq.~\ref{eq:interp} dominates and the free energy approaches that of an ideal gas. When $f_{cv}=1$, the model reproduces the original ANEOS, and the high-temperature heat capacity converges to $3N_0kT$. 

\begin{figure}
    \centering
    \includegraphics[width=\textwidth]{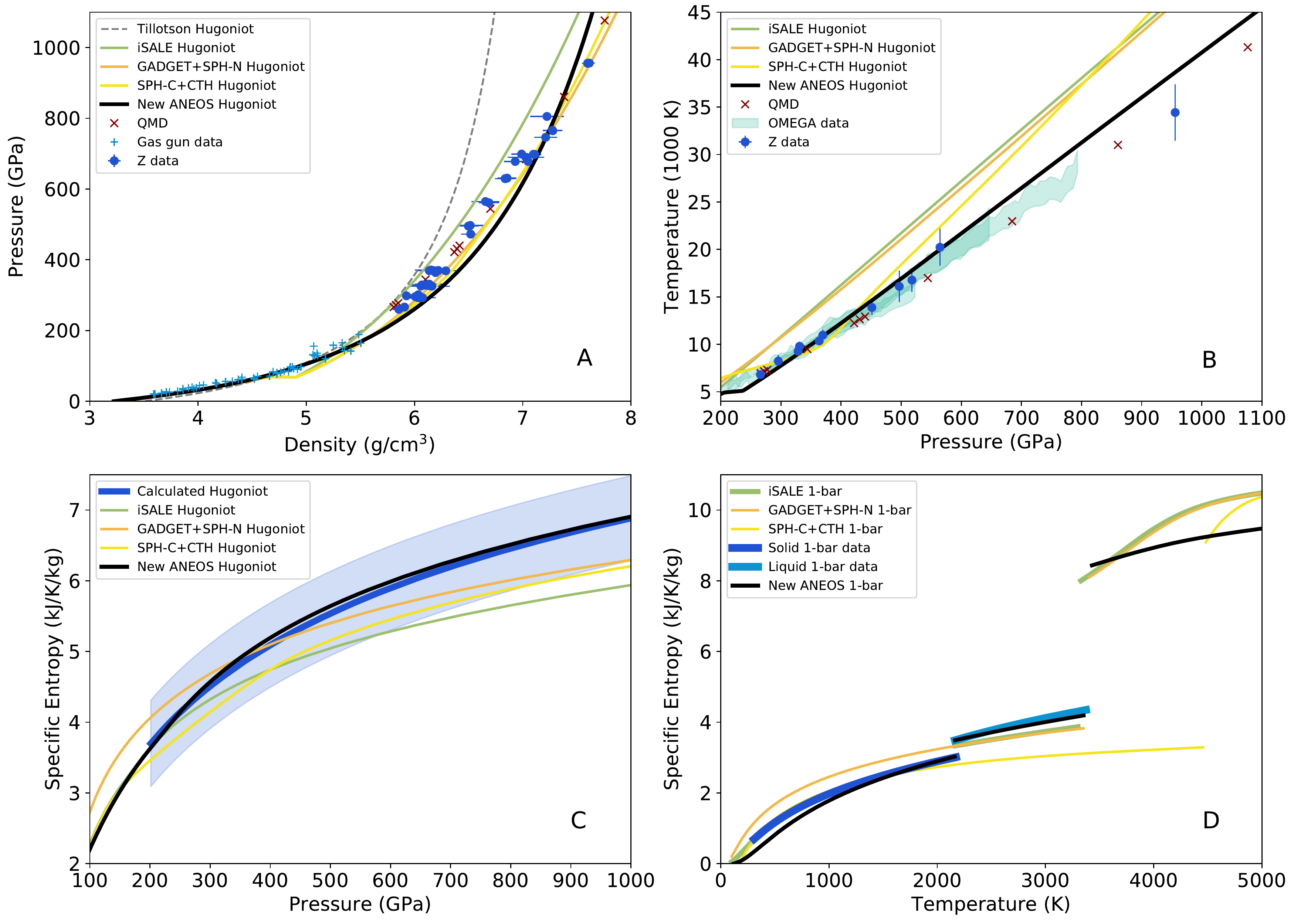}
    \caption{Comparison between data on forsterite, previous versions of ANEOS, and our new model. A. The ANEOS STP Hugoniots used in the iSALE  \cite{collins2014improvements}, GADGET and SPH-N \cite{cuk2012making,Nakajima2014}, SPH-C and CTH codes \cite{canup2013lunar}, and the Tillotson Hugoniot for olivine \cite{marinova2011geophysical}. The STP Hugoniot data are from \cite{mosenfelder2007thermodynamic} (summary of gas gun data) and \cite{Root2018} (Sandia Z machine data and QMD calculations). B. Shock temperatures from the Z machine (1 corrected point and 1 new point described in
    \cite{Davies2019foentropy}), Omega EP laser, and QMD calculations from \cite{Root2018}. C. Specific entropies on the Hugoniot and $1\sigma$ error envelope (shaded region) were calculated by thermodynamic integration \cite{Davies2019foentropy}. D. Experimentally constrained 1-bar specific entropies are from \cite{gillet1991high} (solid) and \cite{thomas2013direct} (liquid), assuming constant liquid specific heat capacity, $c_p$, to 3375~K (the boiling point has not been measured). The jumps correspond to melting (2163~K) and vaporization (models range between 3320 and 4460~K). All data are available in the GitHub repository.}
    \label{fig:hugs}
\end{figure}

\section{REVISED ANEOS MODEL FOR FORSTERITE}

We have used the modified ANEOS model to update the equation of state for forsterite (Mg$_2$SiO$_4$), the magnesium end member of the olivine solid solution series ((Mg,Fe)$_2$SiO$_4$). Olivine is a major mineral in planetary interiors, and forsterite is widely used as a single-component proxy for the silicate mantle of planets. ANEOS has over 40 input parameters, and several different parameter sets for forsterite, sometimes called dunite (an olivine-rich igneous rock), have been used in planetary impact calculations. To date, different hydrocodes have implemented different versions of ANEOS forsterite: e.g., (i) iSALE \cite{collins2014improvements}, (ii)  GADGET \cite{marcus2011role,cuk2012making} and SPH-N \cite{Nakajima2014}, (iii) SPH-C \cite{canup2012forming} and CTH \cite{canup2013lunar,barr2016origin}, and (iv) SPH-B \cite{benz1989origin}. These versions, which also utilized different capabilities in ANEOS, have variable detail in their documentation in terms of the material data that were used to develop the model or in the known limitations of the model. Our model parameter choices and comparisons to material properties are presented in a Jupyter notebook that is available on GitHub along with SESAME and GADGET-format EOS tables.

Our new ANEOS (with $f_{cv}=1.35$) significantly improves the  temperatures and specific entropies on the Hugoniot and the specific entropies of the liquid at 1 bar (Figure~\ref{fig:hugs}). Previous ANEOS models were unable to reproduce the observed shock temperatures because the thermal model converged to the Dulong-Petit limit. These earlier models for forsterite were developed with limited data at the shock pressures of the giant impact regime and no information about the vapor curve. Now, the Hugoniot is measured up to 950~GPa \cite{Root2018}, and the vapor curve is constrained by experiments \cite{davies2019forsterite} and ab initio calculations \cite{mattsson2019critical}. The increased specific heat capacity in the liquid leads to a 1-bar boiling point specific entropy that is slightly larger than in previous EOS models. Note that the ANEOS versions used in GADGET, SPH-N, SPH-C, and CTH do not include a melt curve but do include a high-pressure phase transition. The outcomes of giant impact calculations should be re-evaluated using our revised EOS.

\section{LIMITATIONS OF ANEOS AND TILLOTSON EOS MODELS}

Our model has been developed specifically for high-energy impact processes. Hence, we have focused on the melt curve, liquid region, and vapor curve at the expense of accuracy in the solid phase. As shown in Figure~\ref{fig:hugs}D, the specific entropy of the low-temperature solid is incorrect because the model reference Debye temperature is fitted to match the melt curve and liquid region rather than the low-temperature solid. 

The high-pressure phase transition feature in ANEOS is a preliminary implementation that was never completed. Because the liquid model in ANEOS is currently referenced to the solid, the high-pressure phase transition is also applied to the liquid field. Our new model does not include a high-pressure phase transition because of this undesirable side effect. As a result, the bulk modulus in our model is a compromise to span the solid and liquid fields.

The formulas used for the Gr\"uneisen parameter in ANEOS apply to both the solid and liquid. Previous models have fit to the Gr\"uneisen parameter for the solid. Here, we fit to the Gr\"uneisen parameter for the liquid \cite{thomas2013direct,Davies2019foentropy}; however, the ANEOS formulas are unable to capture the true properties of liquid forsterite, leading to errors at densities below the reference point (the solid at the 1 bar melting point) and densities above about 7.5~g/cm$^3$, corresponding to pressures greater than about 1000~GPa on the shock Hugoniot.  As a result, the curvature of the model Hugoniot is slightly off from the data.

\begin{figure}
    \centering
    \includegraphics[width=\textwidth]{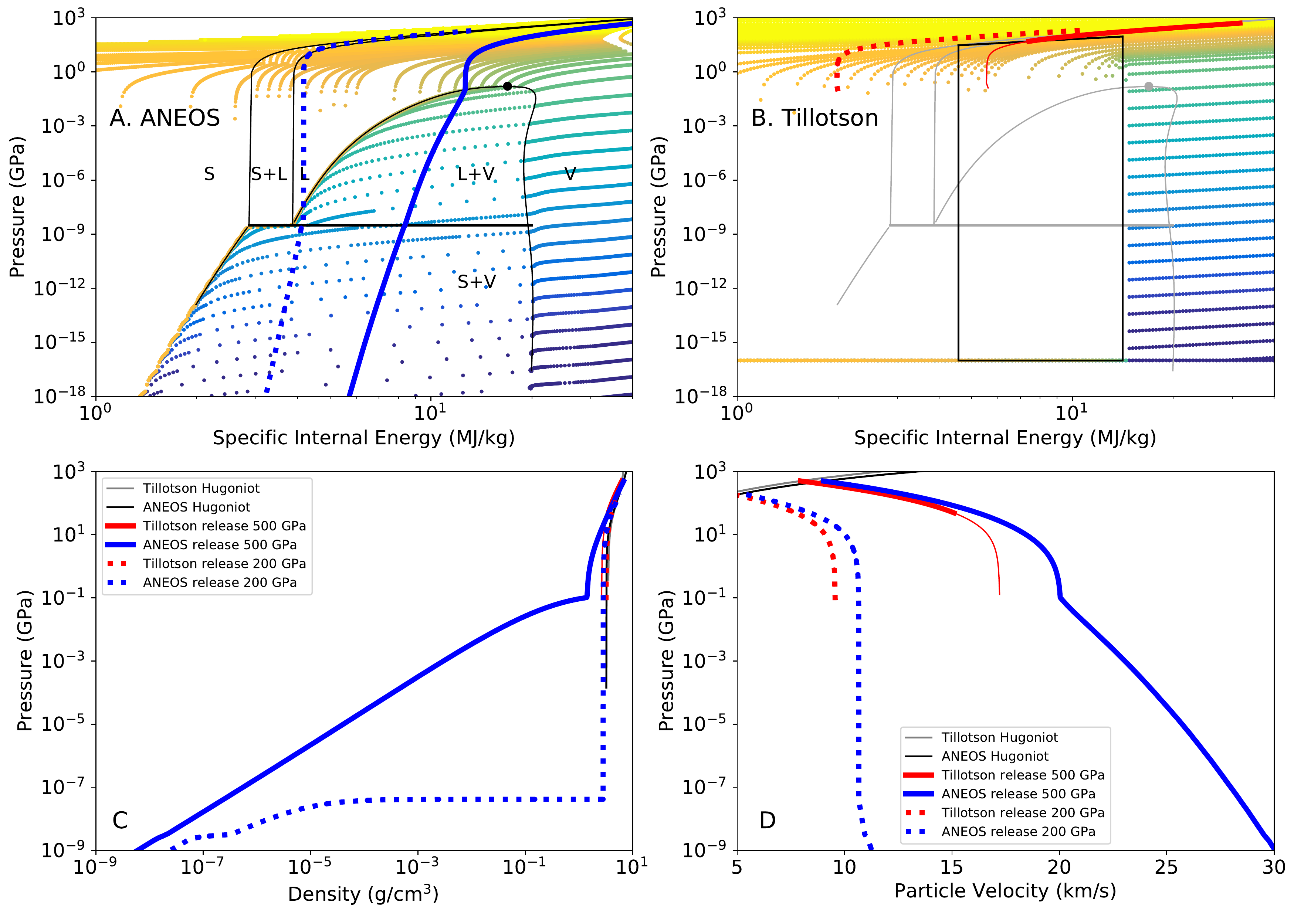}
    \caption{Comparison between our revised ANEOS model for forsterite and a Tillotson model for olivine \cite{marinova2011geophysical}. A. ANEOS isochores (colored dotted lines) shown with the phase boundaries. Each dot is a point in a tabulated density-temperature grid. B. Tillotson isochore dots represent a point in a tabulated density-specific internal energy grid. The black box denotes the interpolated region of the Tillotson EOS. The ANEOS phase boundaries are shown in grey for reference. Thick blue and red lines denote release isentropes from shock pressures of 500~GPa (solid) and 200~GPa (dotted) on the STP Hugoniots. The density decreases rapidly with decompression when the release isentrope intersects the vapor curve (C). Decompression is accompanied by a change in particle (material) velocity (D). The Tillotson release isentropes (red) end at the point where the EOS has no physical definition for the sound speed. The red line for the 500~GPa isentrope becomes thinner when it enters the non-physical interpolated region.}
    \label{fig:tillcomp}
\end{figure}

In Figure~\ref{fig:tillcomp}AB, isochores across the EOS surface illustrate the major differences between our revised ANEOS model for forsterite and Tillotson model for olivine \cite{marinova2011geophysical}. In equilibrium, a decompressing solid/liquid parcel intersects the sublimation/vapor curve in the ANEOS model. Because the Tillotson model does not include phase boundaries, decompression in the cold region returns large negative pressures, which are often reset to a minimum positive value (here, $10^{-16}$~GPa). Similarly, negative sound speeds are also reset to a fixed small value. Thus, there is a large region of the Tillotson phase diagram (low-pressure cold region and interpolated region) where the sound speeds are not physical. In most implementations of Tillotson (e.g., in the iSALE code Dellen version and the SPH code in \cite{hosono2019terrestrial}), the negative pressures in the cold region were used in the interpolation calculation \citep{melosh1989impact}; as a result, most of the interpolated region was reset to minimum pressures.

The sound speed is a critical parameter during decompression from the shocked state. The particle velocity during one-dimensional decompression from an initial shock state with pressure $P_H$ and particle velocity $u_{{\rm p},H}$ to an ambient pressure $P$ is given by the Riemann integral \cite{rice1958compression},
\begin{equation}
u_{\rm p}=u_{{\rm p},H} \pm \int_{P_H}^{P} \frac{dP}{\sqrt{ \left( - \partial P / \partial V \right)_S } } = u_{{\rm p},H} \pm \int_{P_H}^{P} \frac{V}{c_s} dP, \label{eq:release}
\end{equation}
where $c_s$ is the sound speed, $V$ is specific volume, and subscript $S$ in the derivative denotes an isentrope. The $\pm$ indicates the relative orientation of the shock and rarefaction waves, and for illustrative purposes in Figure~\ref{fig:tillcomp}, we present shock-and-release from a downrange free surface (-).

An erroneous EOS leads to errors in the dynamics of the problem. Decompressing material will not only have incorrect density-pressure relationships but also have incorrect material velocities. These physical parameters affect key outcomes of a giant impact such as the amount of mass ejected or emplaced into orbit. The errors are illustrated in Figure~\ref{fig:tillcomp}CB, which show the release isentropes from shock states at 200 and 500~GPa. In the ANEOS model, when the isentrope intersects the vapor dome, the bulk densities decreases and the particle velocities increase. In the Tillotson model, the decompressing material reaches the undefined region of phase space at relatively high pressures. The Tillotson EOS is undefined or interpolated in regions that encompass escaping material from giant impacts and impact-generated circumplanetary disks (e.g., the Moon-forming disk is typically $10^{-4}$ to $10^{-2}$~GPa and partially vaporized at the Roche limit \cite{lock2018origin}). Modeling impact-generated circumplanetary disks requires inclusion of the vapor curve. As discussed in Appendix II of 
\cite{melosh1989impact}, the ANEOS model is superior to the Tillotson model. The Tillotson model is non-physical in the phase space of giant impacts and should not be used for this application. 

\section{Conclusions}

We present an updated version of the ANEOS model for forsterite that is suitable for modeling giant impacts. The model improves the vapor curve and temperatures in the liquid field, but lacks high-pressure solid phases. The commonly-used Tillotson equation of state is not appropriate for giant impacts where vaporization is a key physical process. The revised ANEOS parameter set for forsterite, modifications to the ANEOS code, SESAME and GADGET format EOS tables, and code to generate the figures are available at https://github.com/ststewart/aneos-forsterite-2019. 

\begin{acknowledgments}
This work was supported by the Sandia Z Fundamental Science Program; DOE-NNSA DE-NA0002937 and DE-NA0003842; NASA NNX15AH54G; NASA NNX16AP35H; UC Office of the President LFR-17-449059. Sandia National Laboratories is a multimission laboratory managed and operated by NTESS, LLC, for the DOE-NNSA under contract DE-NA0003525. Prepared by LLNL under contract DE-AC52-07NA27344. Report No.~SAND2019-8839J.
\end{acknowledgments}


%

\end{document}